\documentclass[preprint,aps,prd,nofootinbib,tightenlines]{revtex4}

\usepackage{amsmath, amsfonts, amsthm, amssymb, graphicx}

\def\be{\begin{equation}}
\def\ee{\end{equation}}
\def\ba{\begin{eqnarray}}
\def\ea{\end{eqnarray}}

\def\bdm{\begin{displaymath}}
\def\edm{\end{displaymath}}

\def\bq{\begin{quote}}
\def\eq{\end{quote}}

\def\ltap{\ \raise.3ex\hbox{$<$\kern-.75em\lower1ex\hbox{$\sim$}}\ }
\def\gtap{\ \raise.3ex\hbox{$>$\kern-.75em\lower1ex\hbox{$\sim$}}\ }
\def\gl{\ \raise.5ex\hbox{$>$}\kern-.8em\lower.5ex\hbox{$<$}\ }
\def\roughly#1{\raise.3ex\hbox{$#1$\kern-.75em\lower1ex\hbox{$\sim$}}}

 at 10truept

\newcommand{\beq}{\begin{equation}}
\newcommand{\eeq}{\end{equation}}
\newcommand{\bea}{\begin{eqnarray}}
\newcommand{\eea}{\end{eqnarray}}
\newcommand{\beqa}{\begin{eqnarray}}
\newcommand{\eeqa}{\end{eqnarray}}
\newcommand{\nn}{\nonumber\\}
\newcommand{\ud}{\mathrm{d}}

\def \pd {\partial}

\begin{document}

\title{Multi-galileons, solitons and Derrick's theorem}

\author{Antonio Padilla} 
\email[]{antonio.padilla@nottingham.ac.uk}
\author{Paul M. Saffin} 
\email[]{paul.saffin@nottingham.ac.uk}
\author{Shuang-Yong Zhou} 
\email[]{ppxsyz@nottingham.ac.uk}

\affiliation{School of Physics and Astronomy, 
University of Nottingham, Nottingham NG7 2RD, UK} 

\date{\today}

\begin{abstract}
Galileon models, which were developed in the context of modified gravity, give a class of Lagrangians containing derivative interactions without introducing higher order derivatives in the equations of motion. Here we extend the analysis to an arbitrary number of scalars, and examine the restrictions imposed by an internal symmetry, focussing in particular on SU(N) and SO(N). This therefore extends the possible gradient terms that may be included when constructing topological objects such as sigma model lumps.
\end{abstract}


\maketitle

\section{Introduction}
In an attempt to model the generic features of a particular class of modified gravity scenarios~\cite{dgp, kogan, ruth, me} , Nicolis {\it et al}~\cite{Nicolis:2008in}  recently developed the notion of the {\it galileon}. The galileon is a scalar field, $\pi$, whose dynamics is  described by a Lagrangian that is invariant under Galilean transformations of the form $\pi\longrightarrow\pi+b_\mu x^\mu+c$, where $b_\mu$ and $c$ are constant.  A celebrated example of a galileon theory is derived from the  DGP model~\cite{dgp}. To see this one can compute the boundary effective theory~\cite{Luty:2003vm}, and take a  limit in which an additional scalar degree of freedom decouples from the graviton fluctuations~\cite{Nicolis:2004qq}. This scalar can be identified with the position of the DGP brane in the Minkowski bulk, and its dynamics is described by a Galilean invariant theory in the decoupling limit. The Galilean symmetry is a descendent of Poincar\'e transformations in the bulk. Most of the interesting dynamics occurs in this scalar sector, and one can study a number of phenomenological issues including, self acceleration~\cite{sa}, ghosts~\cite{saghosts}, and the Vainshtein mechanism~\cite{dgp-vainsh}.

Taking the Galilean symmetry as their starting point Nicolis {\it et al}~\cite{Nicolis:2008in} showed that if we also require the field equations to be at most second order in derivatives, then a generic term in the single galileon Lagrangian is of the form
\ba
\eta^{\mu_1}_{\phantom{\mu_1}\![\nu_1}...\eta^{\mu_m}_{\phantom{\mu_m}\!\nu_{m}]}\,\pi
\pd_{\mu_1}\!\pd^{\nu_1} \pi  ... \pd_{\mu_{m}}\!\pd^{\nu_{m}} \pi,
\ea
where $m\leq d$, the number of spacetime dimensions. Many aspects of the single galileon theory have been studied, ranging from phenomenology~\cite{clare}, covariant completions~\cite{covgal}, and  their interpretation in terms of the position of probe DBI branes~\cite{dbigal}. See \cite{david} for early related work.

The extension of this symmetry to two galileon fields has been recently developed~\cite{us} (see also~\cite{pforms}). As with the single galileon this was  motivated by modified gravity scenarios, and in particular, co-dimension two braneworld models~\cite{cod2, koyama, cline, nk, ccpap}. The two galileon fields can be identified with the position of the 3-brane in a six dimensional bulk. Generically there is no reason to expect there to be an internal symmetry linking the two brane bending modes since the brane can fluctuate independently along the two orthogonal directions normal to the brane.  However, one could speculate that the decoupling limit of a rotationally symmetric, co-dimension two brane would correspond to a bi-galileon theory with an internal SO(2) symmetry -- the Galilean symmetry coming from bulk Poincar\'e invariance, and the SO(2) from cylindrical symmetry that remains unbroken by the brane.

Here we take a slightly different view from previous work on galileons in that we do not concern ourselves with gravitational physics. Instead we will consider galileons purely from a field theorist's perspective.  The Lagrangians  describing galileon theories typically include higher order gradient interactions. Higher order gradient terms play a central role in constructing sigma models with stable lumps, such as the Skyrme~\cite{Skyrme:1961vq,Skyrme:1962vh} and baby Skyrme model~\cite{Piette:1994ug}, as they behave differently to the canonical kinetic term under a rescaling of co-ordinates. This allows one to evade Derrick's theorem~\cite{Derrick:1964ww,Manton:2004tk}, and so construct finite energy topological defects.  These solitons have been used to model baryons, so naturally it would be interesting to extend these models, perhaps by introducing new types of gradient interactions to stabilise new configurations, or at least render them static. However, generically, such interaction  terms will introduce the ghost-like instabilities associated with   higher derivative field theories~\cite{Ostro}. The nice thing about galileon theories is that even though higher order gradient terms are present in the Lagrangian, the  equations of motion are no more than second order in derivatives. Thus the ghost problem is avoided, and one is encouraged to extend the Skyrme model in new ways.

In this paper we keep the general structure of the galileon term, but we allow for the scalar to be a multi-component field. If we were not to impose any symmetry between the components this would lead to a huge proliferation of terms rather quickly, but by imposing an internal symmetry we are able to find a manageable set of terms. We start in sections~\ref{sec:SONf} and~\ref{sec:SONa} by constructing terms that utilize real scalars with an SO(N) symmetry, before moving on to a complex scalar realizing SU(N) symmetry in sections~\ref{sec:SUNf} and~\ref{sec:SUNa}. In section~\ref{sec:derrick},  we discuss Derrick's theorem in the context of galileons.  By means of an example, we show explicitly how Derrick's  theorem can be evaded, and one can indeed use the higher order gradient interactions to construct static soliton solutions. This result is in contrast to~\cite{Endlich:2010zj}, where it was argued that galileon interactions could not be used in this way.  The difference here is that we can consider multiple fields constrained to lie  on a non-linear manifold, and so one cannot consider  field deformations orthogonal to the pre-defined target space We will summarize our results in section~\ref{sec:sum}.

\section{Multi-component galileon}

A general multi-galileon theory, in four dimensions, with $N$ \emph{real} degrees of freedom is given by the Lagrangian~\cite{us}
\be \label{lagsonf}
\mathcal{L}_\pi = \sum_{m=1}^5 \alpha^{i_1...i_m}\, \eta^{\phantom{[}\mu_2...\mu_{m}\phantom{]}}_{[\nu_2...\nu_{m}]}
\pi_{i_1}\pd_{\mu_2}\!\pd^{\nu_2}  \pi_{i_2}...\pd_{\mu_{m}}\!\pd^{\nu_{m}} \pi_{i_m},
\ee
where $\{\alpha^{i_1 ... i_m}\}$ are free parameters of the theory and $\eta^{\phantom{[}\mu_1...\mu_{m}}_{[\nu_1...\nu_{m}]}=m!\eta^{\mu_1}_{\phantom{\mu_2}\![\nu_1}...\eta^{\mu_m}_{\phantom{\mu_m}\!\nu_{m}]}$. Note that throughout this paper summation over repeated Lorentz (Greek) and galileon indices (Latin) are understood. Note  further that we {\it define} the $m=1$ term of expression (\ref{lagsonf}) to be $\alpha^{i_1}\pi_{i_1}$. For each galileon term (ie, terms grouped by a $\alpha^{i_1 ... i_m}$), since $\{\mu_k\}$ and $\{\nu_k\}$ are anti-symmetrized respectively, $\pd_{\mu_k}\pd^{\nu_k}$ can be ``moved'' by integration by parts from $\pi_{i_k}$ to $\pi_{i_1}$, thus the galileon indices of $\eta^{\phantom{[}\mu_2...\mu_{m}}_{[\nu_2...\nu_{m}]}\pi_{i_1}\pd_{\mu_2}\pd^{\nu_2} \pi_{i_2}...\pd_{\mu_{m}}\pd^{\nu_{m}} \pi_{i_m}$ are symmetric, which in turn means that we can choose $\alpha^{i_1 ... i_m}$ as symmetric. The total number free parameters in a general $N$-galileon theory is therefore given by
\be
\sum_{ m=1}^{5} \left(\begin{array}{c} N+m-1 \\ m \end{array} \right)=\sum_{m=1}^{5} \frac{(N+m-1)!}{m!(N-1)!}.
\ee 
For example, when $N=3$ there are already 55 free parameters.

Now, if there is an internal rotational symmetry within the $N$-galileon, we will see that the number of free parameters is greatly reduced. 

\subsection{SO(N) fundamental representation} \label{sec:SONf}

Suppose that the $N$-galileon transforms according to the fundamental representation of SO(N), that is, 
\be 
\pi_i\to O_i^{\;\;j}\pi_j,\qquad O_i^{\;\;k}O_j^{\;\;l}\delta_{kl}=\delta_{ij},\qquad i,j,k,l=1,...,N.
\ee
In order for the terms in (\ref{lagsonf}) to be invariant under such a transformation we require the parameters $\{\alpha^{i_1 ... i_m}\}$ to be invariant tensors of the fundamental representation,
\begin{align}
\alpha^{j_1 ... j_m}=\alpha^{i_1 ... i_m}O_{i_1}^{\phantom{i_1}j_1}...O_{i_m}^{\phantom{i_m}j_m}.
\end{align}
Invariant tensors can be constructed from the \emph{primitive invariants} of the group in the corresponding representation; the primitive invariants of the fundamental representation of SO(N) are the Kronecker delta and the Levi-Civita symbol~\cite{gpbk}
\be
 \delta^{ij},\qquad \epsilon^{i_1... i_N}.
\ee
Since $\alpha^{i_1 ... i_m}$ is symmetric, we can ignore the anti-symmetric Levi-Civita symbol and construct $\alpha^{i_1 ... i_m}$ solely from $\delta^{ij}$. Then it is clear that, up to 5th order, the non-zero free parameters are $\alpha^{ij}$ and $\alpha^{ijkl}$, which can be parametrised as
\ba
 \alpha^{ij}=\alpha_{(2)}\delta^{ij},\qquad              \alpha^{ijkl}=\alpha_{(4)} \delta^{(ij} \delta^{kl)}.
\ea
So the most general SO(N) fundamental galileon Lagrangian is given by
\ba
\mathcal{L}_\pi&=&\alpha_{(2)}\,\pi^i\Box\pi_i
                 +\alpha_{(4)}\, \eta^{\phantom{[}\rho\mu\lambda}_{[\sigma\nu\tau]}\pi^i\pd_{\rho}\pd^{\sigma}\pi_i \pd_{\mu}\pd^{\nu}\pi^j
                                                                                        \pd_{\lambda} \pd^{\tau}\pi_j\\
              &=& \alpha_{(2)}\,\underline\pi^T\Box\underline\pi 
               +\alpha_{(4)}\, \eta^{\phantom{[}\rho\mu\lambda}_{[\sigma\nu\tau]}
                        \left(\underline\pi^T\pd_{\rho}\pd^{\sigma}\underline\pi\right) 
                        \left(\pd_{\mu}\pd^{\nu}\underline\pi^T\pd_{\lambda} \pd^{\tau}\underline\pi\right) .
\ea

\subsection{SO(N) adjoint representation} \label{sec:SONa}

We have seen that constructing a general SO(N) galileon Lagrangian for any $N\geqslant 2$ in the fundamental representation is straightforward, but it is not necessarily a simple procedure for scalars that transform in representations other than the fundamental. Another possibility we are particularly interested in is the adjoint representation, which may be conveniently described by $N\times N$ anti-symmetric matrices, $\Pi_{ij}=-\Pi_{ji}$, transforming as
\be
\Pi \to O \Pi O^{T},\qquad \mathrm{or}~~~ \Pi_{ij}\to O_i^{\;\;k}\Pi_{kl}O_j^{\;\;l},
\ee
where $O=(O_{i}^{\phantom{i}j})_{N\times N}$ again is a SO(N) defining matrix. Therefore a general SO(N) Lagrangian with a galileon transforming in the adjoint representation may be written as
\be
\label{eq:adLag}
\mathcal{L}_{\Pi}=\sum_{m=1}^5\beta^{i_1 j_1,...,i_m j_m}\, \eta^{\phantom{[}...\phantom{]}}_{[...]}\Pi_{i_1 j_1}
\pd\pd \Pi_{i_2 j_2}...\pd\pd \Pi_{i_m j_m}.
\ee
Note that for simplicity we suppress the Lorentz indices in this section. Treating the pair ($i_kj_k$) as ``one index'', $\beta^{i_1 j_1,..., i_m j_m}$ can be considered as adjoint invariant tensors. Since $\eta^{\phantom{[}...\phantom{]}}_{[...]}\pd\pd \Pi_{i_2 j_2}...\pd\pd \Pi_{i_m j_m}$ is symmetric in the galileon pairs, $\beta^{i_1 j_1,..., i_m j_m}$ can then be chosen as symmetric in the galileon index pairs (within each pair, exchanging $i_k$ and $j_k$ is of course anti-symmetric). To determine all possible $\beta^{i_1 j_1,...,i_m j_m}$, we would like to find all the \emph{symmetric} primitive symmetric invariants up to 5th order (in terms of pairs) in this representation, and then construct $\beta^{i_1 j_1,...,i_m j_m}$ using these primitive invariants. As an aside, since the indices (not pairs) still transform under fundamental SO(N) matrices, we expect, as we shall see shortly, that all the primitive invariants in the adjoint representation can actually be constructed from $\delta^{ij}$ and $\epsilon^{i_1... i_N}$.

Before the construction, however, we would like to clear up some special cases of the low rank groups. $SO(2)$ is Abelian, and not surprisingly the adjoint representation of $SO(2)$ has only one independent component, thus the adjoint $SO(2)$ galileon reduces to the single galileon case. The adjoint of $SO(3)$ is equivalent to the fundamental, so the adjoint of $SO(3)$ offers no new model. $SO(4)$ is not a simple Lie group and is locally isomorphic to $SU(2)\times SU(2)$, so the adjoint of $SO(4)$ is reducible. Nevertheless, in terms of constructing Lagrangian singlets, we shall see that it goes along similar lines as some other simple $SO(2l)$. We shall ignore $SO(2)$ and $SO(3)$ and consider SO(N) with $N\geqslant4$ in the following.

For a simple Lie group $G$, it is well known that there are $l(=\mathrm{rank}(G))$ symmetric primitive invariants in the adjoint \cite{gpinv}. For simple $SO(2l+1)$ and $SO(2l)$, $l$ and $l-1$ of them  are given by the first $l$ and $l-1$ \emph{even} order (in terms of pairs) of the $k$-family symmetric invariant tensors respectively, defined by the symmetrized trace
\begin{align}
k^{i_1j_1,...i_mj_m}=\frac1{m!}\mathrm{sTr}(T^{i_1j_1}...\;T^{i_mj_m}),
\end{align}
where $T^{i_kj_k}$ are the generators of the fundamental representation of SO(N), conventionally chosen as
\be
(T^{kl})_i^{\;\;j}=-i\left(\delta^k_i\delta^{lj}-\delta^l_i\delta^{kj}\right).
\ee
By noting that the $\Pi$ terms in the Lagrangian (\ref{eq:adLag}) are already symmetric under interchanging the pairs ${(i_kj_k)}$, the symmetrizing of the trace is done for us, leading to the most general Lagrangian for these cases to be
\begin{align}
\mathcal{L}_\Pi=&~\beta_{(2)}\mathrm{Tr}(\Pi\Box\Pi)
+ \beta_{(4)}\eta^{\phantom{[}...\phantom{]}}_{[...]}\mathrm{Tr}(\Pi
\pd\pd \Pi)\mathrm{Tr}(\pd\pd \Pi\pd\pd \Pi)
    + \beta_{(4)}' \eta^{\phantom{[}...\phantom{]}}_{[...]}\mathrm{Tr}(\Pi
\pd\pd \Pi\pd\pd \Pi\pd\pd \Pi).
\end{align}
For $SO(2l)$ there is one more primitive invariant, associated to the Pfaffian, which for $l=2,3,4,5$ allows us to construct terms that may appear in the Lagrangian; for larger $l$ they lead to terms with more than five $\Pi$s, and so cannot contribute. The invariants are
\begin{align}
l&=2:&&\epsilon^{i_1j_1i_2j_2},\\
l&=3:&&\epsilon^{i_1j_1...i_3j_3},\\
l&=4:&&\epsilon^{i_1j_1...i_4j_4},\\
l&=5:&&\epsilon^{i_1j_1...i_5j_5},
\end{align}
which we note are symmetric in moving pairs of indices. The existence of these invariants is related to the fact that the determinant of an anti-symmetric matrix $(M_{ij})_{2N\times 2N}$ can be expressed as the square of the Pfaffian,
\begin{align}
\mathrm{Pf}(M)&=\sqrt{\mathrm{det}(M)}\nn
        &=\frac{1}{2^N N!}\epsilon^{i_1j_1...i_Nj_N}M_{i_1j_1}...M_{i_Nj_N}.
\end{align}
These invariants give rise to additional terms for each of the four groups ($SO(2l)$, $l=2,3,4,5$) respectively:
\begin{align}
\mathcal{L}_{\Pi}^4  &=\eta_4 \;\epsilon(\Pi\Box \Pi)
+\eta'_4 \eta^{\phantom{[}...\phantom{]}}_{[...]}\;\epsilon(\Pi\pd\pd \Pi)  \mathrm{Tr}(\pd\pd\Pi\pd\pd\Pi),\\
\mathcal{L}_{\Pi}^6  &=\eta_6 \eta^{\phantom{[}...\phantom{]}}_{[...]}\;\epsilon(\Pi
\pd\pd \Pi\pd\pd\Pi)
+\eta'_6 \eta^{\phantom{[}...\phantom{]}}_{[...]}\;\epsilon(\Pi
\pd\pd \Pi\pd\pd\Pi)\mathrm{Tr}(\pd\pd\Pi\pd\pd\Pi),\\
\mathcal{L}_{\Pi}^8   &=\eta_8 \eta^{\phantom{[}...\phantom{]}}_{[...]}\;\epsilon(\Pi
\pd\pd \Pi\pd\pd\Pi\pd\pd\Pi),\\
\mathcal{L}_{\Pi}^{10}  &=\eta_{10} \eta^{\phantom{[}...\phantom{]}}_{[...]}\;\epsilon(\Pi
\pd\pd \Pi\pd\pd\Pi\pd\pd\Pi\pd\pd\Pi),
\end{align}
with the definition 
\be
\epsilon(\Pi\pd\pd \Pi
...\pd\pd \Pi)=\epsilon^{i_1j_1...i_mj_m}\Pi_{i_1j_1}
\pd\pd \Pi_{i_2 j_2}...\pd\pd \Pi_{i_m j_m}.
\ee
Note that for $SO(4)$ one might have expected an additional singlet term proportional to 
\be
\eta^{\phantom{[}...\phantom{]}}_{[...]}\epsilon(\Pi
\pd\pd \Pi)\;\epsilon(\pd\pd\Pi \pd\pd \Pi)  .
\ee
However, by noting the relation
\begin{align}
\epsilon^{i_1j_1i_2j_2}\epsilon_{i_3j_3i_4j_4}=
4!\delta^{[i_1}_{i_3}\delta^{j_1}_{j_3}\delta^{i_2}_{i_4}\delta^{j_2]}_{j_4},
\end{align}
we see this term degenerates to a linear combination of 
$\eta^{\phantom{[}...\phantom{]}}_{[...]}\mathrm{Tr}(\Pi\pd\pd \Pi)\mathrm{Tr}(\pd\pd \Pi\pd\pd \Pi)$ and 
$\eta^{\phantom{[}...\phantom{]}}_{[...]}\mathrm{Tr}(\Pi\pd\pd \Pi\pd\pd \Pi\pd\pd \Pi)$.

\subsection{SU(N) fundamental representation} \label{sec:SUNf}

In this section, we promote the $N$ galileons to be $N$ complex fields $\phi_i$ and impose SU(N) symmetry within the complex $N$-galileon $\phi=(\phi_1,...,\phi_N)$. The tensor product of two irreducible representations can produce a singlet only if they are conjugate to each other. For SO(N), the complex conjugate of the fundamental is equivalent to the fundamental itself, thus we could build singlets merely from the fundamental representation. For SU(N), this is generally not the case, so we include fields that transform with the complex conjugate representation $\phi^*$ in building the SU(N) fundamental galileon. Since the conjugate representation of SU(N) is generally inequivalent to the fundamental representation, we would have to differentiate the indices of the two representations carefully. A simple way to do so is representing the fundamental field with a lower index $\phi_i$ and representing the conjugate field with a upper index $\phi^i=(\phi_i)^*$, such that they transform as
\ba
\phi_i\to U_i^{\;\;j}\phi_j,\qquad\phi^{i}\to U^i_{\;\;j}\phi^{j},\qquad U_i^{\;\;j}U^k_{\;\;j}=\delta^k_i.
\ea

A general SU(N) fundamental Lagrangian would be given by
\begin{align} \label{lagsunf}
\mathcal{L}_\phi=\sum_{m=1}^5\alpha^{i_1...i_m}\eta^{\phantom{[}\mu_2...\mu_{m}\phantom{]}}_{[\nu_2...\nu_{m}]}
           \phi_{i_1}\!\pd_{\mu_2}\!\pd^{\nu_2} \phi_{i_2}...\pd_{\mu_{m}}\!\pd^{\nu_{m}} \phi_{i_m}
    +\mathrm{terms~with~conjugate~reps},
\end{align}
where ``terms~with~conjugate~reps'' represents terms with all possible combinations of the fundamental and conjugate representations, ie, all possible terms with upper ${i_k}$ in $\phi$ (correspondingly lower ${i_k}$ in $\alpha^{...}_{\;\;\;\;...}$). 
The primitive invariants of SU(N) fundamental and conjugate representations are given by the Kronecker delta and the Levi-Civita symbols~\cite{gpbk}
\be
 \delta^{i}_{j},\qquad \epsilon^{i_1... i_N},\qquad \epsilon_{i_1... i_N},
\ee 
and, as with the SO(N) case, we can construct our invariant tensors $\alpha^{...}_{\;\;\;\;...}$ out of them. Since $\alpha^{...}_{\;\;\;\;...}$ are totally symmetric, i.e.,
\ba
 \alpha^{i_1i_2...}_{\;\;\;\;\;\;\;\;\;\;j_1j_2...}
=\alpha^{i_2i_1...}_{\;\;\;\;\;\;\;\;\;\;j_1j_2...}
=\alpha^{i_1\;\;i_2...}_{\;\;j_1\;\;\;\;\;\;j_2...}=...,
\ea
we can ignore the anti-symmetric Levi-Civita symbols and construct $\alpha^{...}_{\;\;\;\;...}$ solely from $\delta^{i}_{j}$. 
Therefore, up to 5th order, the only non-zero free parameters are
\ba
\alpha^{i}_{\phantom{i}j}=\alpha_{(1,1)}\delta^{i}_{j},\qquad
\alpha^{ik}_{\;\;\;\;jl}=\alpha_{(2,2)} \delta^{(i}_{\;j} \delta^{k)}_{l},
\ea
and then the SU(N) fundamental galileon Lagrangian is given by
\ba
\mathcal{L}_\phi&=&\alpha_{(1,1)}\,\phi^i\Box\phi_i 
+\alpha_{(2,2)}\, \eta^{\phantom{[}\rho\mu\lambda}_{[\sigma\nu\tau]}\phi^i\pd_{\rho}\pd^{\sigma}\phi_i \pd_{\mu}\pd^{\nu}\phi^j
\pd_{\lambda} \pd^{\tau}\phi_j\\
              &=&\alpha_{(1,1)}\,\phi^{\dagger}\Box\phi 
      +\alpha_{(2,2)}\, \eta^{\phantom{[}\rho\mu\lambda}_{[\sigma\nu\tau]}\left(\phi^{\dagger}\pd_{\rho}\pd^{\sigma}\phi\right)\left( \pd_{\mu}\pd^{\nu}\phi^{\dagger}
\pd_{\lambda} \pd^{\tau}\phi\right).
\ea

\subsection{SU(N) Adjoint Representation} \label{sec:SUNa}

We now turn to construct the most general Lagrangian for a galileon transforming under the SU(N) adjoint representation. Similar to that of the SO(N) adjoint, it is convenient to use $N\times N$ traceless hermitian matrices $\Phi=(\Phi_{i}^{\phantom{i}j})_{N\times N}$ to represent the adjoint SU(N) galileon, which transforms under a SU(N) adjoint action as
\be
\Phi \to U \Phi U^{\dagger}
\ee
where $U=(U_{i}^{\phantom{i}j})_{N\times N}$ again is a SU(N) defining matrix. Therefore a general SU(N) adjoint Lagrangian may be written as
\be
\mathcal{L}_{\Phi}=\sum_{m=1}^5\beta^{i_1 \phantom{j_1,}...\phantom{,}i_m \phantom{j_m}}_{\phantom{i_1} j_1,...,\phantom{i_m} j_m} \eta^{\phantom{[}...\phantom{]}}_{[...]}\Phi_{i_1 \phantom{j_1}}^{\phantom{i_1}j_1}  \pd\pd \Phi_{i_2 \phantom{j_2}}^{\phantom{i_2}j_2}...\pd\pd \Phi_{i_m \phantom{j_m}}^{\phantom{i_m}j_m}.
\ee
Again we suppress the Lorentz indices in this section. Treating the pair ($i_kj_k$) as ``one index'', $\beta^{i_1 \phantom{j_1,}...\phantom{,}i_m \phantom{j_m}}_{\phantom{i_1} j_1,...,\phantom{i_m} j_m}$ can be considered as symmetric invariant tensors (within the pairs, $i_k$ and $j_k$ are hermitian). So the task again is to find all the symmetric primitive invariants up to 5th order (in terms of pairs) in this representation and then construct $\beta^{i_1 \phantom{j_1,}...\phantom{,}i_m \phantom{j_m}}_{\phantom{i_1} j_1,...,\phantom{i_m} j_m}$ from these primitive invariants. Since $\beta^{i_1 \phantom{j_1,}...\phantom{,}i_m \phantom{j_m}}_{\phantom{i_1} j_1,...,\phantom{i_m} j_m}$ always have upper and lower indices and products of $\epsilon^{i_1... i_N}$ and $\epsilon_{j_1... j_N}$ can be reduced to product of $\delta^{i}_{j}$, we would expect that all these primitive invariants can be solely constructed from $\delta^{i}_{j}$.

SU(N) is a rank $N-1$ simple group, so there are $N-1$ symmetric primitive invariants, which may also take to be given by the symmetric traces of the adjoint generators~\cite{gpinv}
\be \label{sunpi}
k^{i_1 \phantom{j_1,}...\phantom{,}i_m \phantom{j_m}}_{\phantom{i_1} j_1,...,\phantom{i_m} j_m}~~~m=2,...,N .
\ee
The adjoint generators of SU(N), however, are more complicated than those of SO(N), including $N(N-1)/2$ the anti-symmetric generators, essentially the same as that of SO(N), as well as $(N+2)(N-1)/2$ symmetric ones. Nevertheless, since the only building block we can use is $\delta^{i}_{j}$, it is not difficult to construct a new family of invariant tensors suitable for constructing the $\beta^{i_1 \phantom{j_1,}...\phantom{,}i_m \phantom{j_m}}_{\phantom{i_1} j_1,...,\phantom{i_m} j_m}$: 
\ba
p^{i_1 \phantom{j_1,}...\phantom{,}i_m \phantom{j_m}}_{\phantom{i_1} j_1,...,\phantom{i_m} j_m}
=\delta^{(i_1}_{j_1}...\delta^{i_m)}_{j_m}.
\ea
The first $N-1$ ($m=2,...,N$) of $p^{i_1 \phantom{j_1,}...\phantom{,}i_m \phantom{j_m}}_{\phantom{i_1} j_1,...,\phantom{i_m} j_m}$ are primitive. Indeed, $p^{i_1 \phantom{j_1,}...\phantom{,}i_m \phantom{j_m}}_{\phantom{i_1} j_1,...,\phantom{i_m} j_m}$ gives rise to the $m$-th order trace $\eta^{\phantom{[}...\phantom{]}}_{[...]}\mathrm{Tr}(\Phi...\pd\pd\Phi)$. 
Therefore, for SU(N) with $N\geqslant5$, the most general Lagrangian with galileon fields transforming under the adjoint of SU(N) is
\begin{align}
\label{eq:sunAdjoint}
\mathcal{L}_\Phi=&~\gamma_2\mathrm{Tr}(\Phi\Box\Phi)+\gamma_3 \eta^{\phantom{[}...\phantom{]}}_{[...]}\mathrm{Tr}(\Phi
\pd\pd \Phi\pd\pd \Phi)   \nn
&~+ \gamma_4 \eta^{\phantom{[}...\phantom{]}}_{[...]}\mathrm{Tr}(\Phi
\pd\pd \Phi\pd\pd \Phi\pd\pd \Phi)
+ \gamma'_4 \eta^{\phantom{[}...\phantom{]}}_{[...]}\mathrm{Tr}(\Phi
\pd\pd \Phi)\mathrm{Tr}(\pd\pd \Phi\pd\pd \Phi)   \nn
&~+\gamma_5 \eta^{\phantom{[}...\phantom{]}}_{[...]}\mathrm{Tr}(\Phi
\pd\pd \Phi\pd\pd \Phi\pd\pd \Phi\pd\pd \Phi)
+\gamma'_5 \eta^{\phantom{[}...\phantom{]}}_{[...]}\mathrm{Tr}(\Phi
\pd\pd \Phi)\mathrm{Tr}(\pd\pd\Phi \pd\pd \Phi\pd\pd \Phi)  .
\end{align}

For lower rank groups we have degeneracies, as a result of the following identity for an arbitrary tensor $ A^{i_1 \phantom{j_1}...\phantom{}~i_m \phantom{j_m}}_{\phantom{i_1} j_1...\phantom{i_m} j_m} $
\be
 \delta^{i_1}_{[j_1}...\delta^{i_{m}}_{j_{m}]} A^{i_1 \phantom{j_1}...\phantom{}~i_m \phantom{j_m}}_{\phantom{i_1} j_1...\phantom{i_m} j_m} =0 \qquad \textrm{for $m>N$} ,
\ee
where the group indices run from $1$ to $N$. Using this result, we can derive the following identities:
\be
 \int\!\! d^4 x  ~\eta^{\phantom{[}...\phantom{]}}_{[...]}\mathrm{Tr}(\Phi(\pd\pd \Phi)^4)
= \int\!\! d^4 x~ \frac56\eta^{\phantom{[}...\phantom{]}}_{[...]}\mathrm{Tr}(\Phi\pd\pd \Phi)\mathrm{Tr}((\pd\pd\Phi)^3) 
\ee
for $N=2, 3, 4$.
\be
\int\!\! d^4 x  ~\eta^{\phantom{[}...\phantom{]}}_{[...]}\mathrm{Tr}(\Phi(\pd\pd \Phi)^3)
=\int\!\! d^4 x  ~\frac12\eta^{\phantom{[}...\phantom{]}}_{[...]}\mathrm{Tr}(\Phi\pd\pd \Phi)\mathrm{Tr}((\pd\pd\Phi)^2)
\ee
for $N=2, 3$ only.  
\be
\int\!\! d^4 x  ~\eta^{\phantom{[}...\phantom{]}}_{[...]}\mathrm{Tr}(\Phi(\pd\pd \Phi)^2)=0
\ee
for $N=2$ only. 

Therefore, for $SU(4)$, the $\gamma'_5$ term degenerates with the $\gamma_5$ term, thus can be dropped. For $SU(3)$, we can further drop the $\gamma'_4$ term. For $SU(2)$, we can further drop the remaining odd order terms, left with only the $\gamma_2$ and $\gamma_4$ terms. This is what one may expect, since $SU(2)$ is locally isomorphic to $SO(3)$.

\section{Derrick's theorem and examples} \label{sec:derrick}

In carrying out this work we have largely been motivated by the possibility of using these terms to find static topological defects, and one of the key ideas to understanding this is Derrick's theorem \cite{Derrick:1964ww,Manton:2004tk}. Derrick's theorem uses the fact that static solutions extremize the energy functional, and so by considering rather simple deformations of a putative static solution, one is able to make statements about their (non)existence. An extension of Derrick's theorem to the single galileon case was considered in \cite{Endlich:2010zj}, where they concluded that such terms cannot stabilize solitons, here we see how one can get around this conclusion. The key to understanding the statement in \cite{Endlich:2010zj} is that they were thinking of the galileon as a linear sigma model and so considered the deformed field configuration
\ba
\label{eq:derrick}
\phi_{\lambda,\omega}(\underline x)&=&\omega\phi_0(\lambda\underline x),
\ea
where $\lambda$ and $\omega$ are constants, and $\phi_0$ is the putative soliton solution. It is then a straightforward calculation to show that the term with $n$ scalars contributes the following to the energy functional,
\ba
E_n(\lambda,\omega)&=&\frac{1}{\lambda^{d+2}}(\lambda^2\omega)^nE_n^{(0)}.
\ea
It is now clear that the energy functional is not stationary under variations of $\omega$. Indeed, if we reduce $\omega$ {\it all} the terms in the energy functional will decrease. However, in non-linear sigma models, such as the Skyrme and baby-Skyrme models, we force the scalar field to live on some particular manifold, $S^3$ and $S^2$ in these respective cases. This means that we are simply not allowed to do the full rescaling put forward in (\ref{eq:derrick}), and we are restricted to $\omega=1$, giving an energy for the rescaled solution (in four spacetime dimensions) of
\ba
E(\lambda)&=&\lambda^{-1}E_2^{(0)}+\lambda E_3^{(0)}+\lambda^{3}E_4^{(0)}+\lambda^{5}E_5^{(0)}.
\ea
This has a minimum as we vary $\lambda$, so solitons in non-linear sigma models are not ruled out. Note that these statements do not rely on any particular representation, or symmetry group. In what follows we give an explicit example of a static soliton using the fundamental representation of SO(4).

\begin{figure}
\includegraphics[height=2in,width=2.9in]{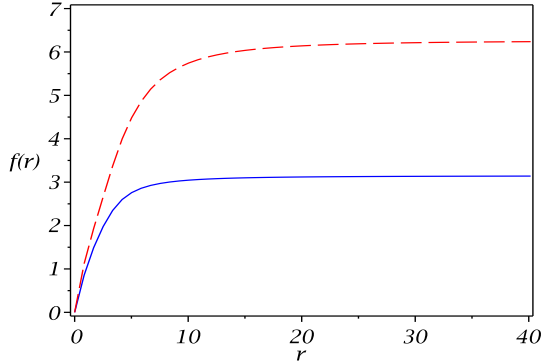}
\caption{Static solitonic solutions for winding number $n=1$ (solid-blue line) and $n=2$ (dashed-red line). The spherically symmetic ansatz (\ref{ansatz}) is used. The energy of them are $E^{(1)}\simeq 101.8$ and $E^{(2)}\simeq 462.3$ respectively.} \label{figfr}
\end{figure}

The SO(4) non-linear sigma model Lagrangian is given by 
\be 
{\cal L}=-\frac{1}{2} \pd^\mu \underline \pi^T
\pd_\mu \underline \pi-\eta^{\phantom{[}\rho\mu\lambda}_{[\sigma\nu\tau]}
                        \left(\underline\pi^T\pd_{\rho}\pd^{\sigma}\underline\pi\right) 
                        \left(\pd_{\mu}\pd^{\nu}\underline\pi^T\pd_{\lambda} \pd^{\tau}\underline\pi\right), 
\ee
where we have made a choice of units such that the dimensionful parameter in front of the quartic piece is absorbed into a co-ordinate scaling. The corresponding Hamiltonian for this theory is given by
 \begin{align}
H= \int d^3 x &~\frac{1}{2} \dot{\underline \pi}^T \dot{ \underline \pi} - 2 \delta^{\phantom{[}ab}_{[de]}\left[(\dot{\underline \pi}^T \dot{ \underline \pi})(\pd_a\pd^ d  \underline\pi^T\pd_b \pd^e \underline \pi)+2(\dot{ \underline \pi}^T \pd_a\pd^ d  \underline\pi)(\dot{ \underline \pi}^T \pd_b\pd^e \underline \pi)\right] \nn
& + \frac12 \pd_a\underline\pi^T\pd^a\underline\pi+\;\delta^{\phantom{[}abc}_{[def]}
                        \left(\underline\pi^T\pd_{a}\pd^{d}\underline\pi\right) 
                        \left(\pd_{b}\pd^{e}\underline\pi^T\pd_{c} \pd^{f}\underline\pi\right),
\end{align}
where $a,\;b,\;c,...$ refer to spatial indices, and `` dot" corresponds to differentiation with respect to time.  We now define our static ansatz for the four-component field
\be \label{ansatz}
\underline{\pi}^T=\left(\frac{x}{r}\sin f(r),\frac{y}{r}\sin f(r),\frac{z}{r}\sin f(r),\cos f(r)\right).
\ee
We will be interested in solutions with integer winding number $n$, satisfying $f(r) \to \pi n$, as $r \to \infty$. Substituting this  into the Hamiltonian yields the following energy functional
\begin{align}
E[f]=4\pi \! \int\!\! \ud r  &~ 2(f')^4-\frac{8\sin f \cos f}{r}(f')^3
    +\frac{r^4+24\sin^2 f-16\sin^4 f}{2r^2}(f')^2 \nn
&  +\frac{r^4 \sin^2 f- 4\sin^4 f}{r^4}  ,
\end{align}
where we have performed some integration by parts to eliminate $f''(r)$. Varying the energy functional to get static solutions,  $\delta E/\delta f=0$ , we obtain the smooth profiles for winding numbers $n=1$ and $n=2$ respectively, as shown in Fig.~\ref{figfr}. The energy of these two profiles is finite and, by explicit integration, take the values $E^{(1)}\simeq 101.8$ and $E^{(2)}\simeq 462.3$ in these units. Note that since $E^{(2)}>2E^{(1)}$, the profile with winding number two is unstable and can decay to two solitons of winding number one.

Finding static solutions with a topological winding number is one thing, but checking their stability is another. For example, although topological textures have a winding number they are certainly not stable, see for example \cite{Turok:1990gw}. So, while the above solutions are extrema of the Hamiltonian, they are not necessarily minima, and a full perturbative analysis of this system is left for future work.

\section{Summary} \label{sec:sum}

In this paper we have expanded on the bi-galileon field theory studied in~\cite{us}, which itself was a generalization of the galileon field theory examined in~\cite{Nicolis:2008in}. Motivated by certain braneworld setups and also to avoid rapid proliferation of possible multi-galileon terms, we have imposed internal symmetries for the multi-galileon field. There are clearly many different possibilities for the internal transformation properties. In this paper we have focused on four cases, namely the fundamental and adjoint representations of $SO(N)$ and $SU(N)$. We have written down the most general galileon theories with these symmetries, and shown that they contain a small number of terms. Incidentally, one can check the results using group theory arguments that we have found the correct number of terms~\cite{liepack}, the point being that these terms correspond to singlets in the symmetric products of the representations. For example, the adjoint representation of $SU(5)$ is the {\bf 24}; so, if we want to know how many $(\Phi)^4$ terms there should be, we need to consider
\ba
{\bf 24}\times {\bf 24}\times {\bf 24}\times {\bf 24}={\bf 1}\oplus{\bf 1}\oplus{\bf 24}\oplus{\bf 24}\oplus{\bf 24}\oplus...,
\ea
ie, there are two singlets. This agrees with the number of quartic $\Phi$ terms in (\ref{eq:sunAdjoint}), namely two. We have only explicitly shown results for symmetric multi-galileon in four spacetime dimensions, but it is clear how our techniques can be extended to any specific number of dimensions.

In galileon models, as well as the canonical kinetic term, one finds terms with more than two derivatives. Such terms behave differently under a rescaling of coordinates, $x^\mu\to \lambda x^\mu$, meaning that Derrick's theorem allows for the possibility of stationary finite energy solitons. Higher order terms in the action are not unusual in the theory of solitons, indeed, they are {\it required}  in the Skyrme model. Here we are modifying the Skyrme model, constructing static solitons using these galileon terms.

To have an explicit example, we have constructed hedgehog solutions for the $SO(4)$ nonlinear sigma multi-galileon confined on $S^3$. The static solutions we obtained possess topological charges, an encouraging sign for the stability of a soliton, but certainly not a proof, so we would like to conclude this paper by saying a word or two about stability. For a non-linear sigma model constructed out of a single galileon, it has been argued that zero mode excitations inevitably suffer from ghost-like instabilities~\cite{Endlich:2010zj}. Those arguments do not necessarily apply here for the following reason.  The act of constraining  $\underline\pi$ to lie on an $S^3$ is equivalent to adding a  constraint term $\lambda(|\underline\pi|^2-1)$ to the Lagrangian.  Perturbatively this gives rise to  a ``mass term'' $\bar \lambda (\delta \underline\pi)^2$, where the background Lagrange multiplier $\bar \lambda$ depends on the $\underline\pi$ profile.  Numerically, we can confirm that $\bar \lambda$ is always negative; thus we come back to the same structure as kinks and skyrmions, where the zero-mode instability arguments of~\cite{Endlich:2010zj} do not apply. Of course, to confirm the stability, extensive numerical studies are required that are beyond the scope of this paper, and which constitutes our future work.

~\\

{\bf Acknowledgements}:~We would like to thank Zhong-Qi Ma and Stanley Deser for useful discussions. AP was funded by a Royal Society University Research Fellowship and SYZ by a CSRS studentship.

\end{document}